\documentclass[superscriptaddress,nofootinbib,twocolumn]{revtex4-1}

\usepackage{hyperref}
\usepackage{bbm}
\usepackage{amsfonts}
\usepackage{mathrsfs}

\usepackage{amsmath,amssymb}

\usepackage{epsfig}
\usepackage{graphicx}               
\usepackage{url}
\usepackage{hyperref}
\usepackage{float}
\usepackage{pstricks}
\usepackage{color}
\usepackage{multirow}
\makeatletter
\def\simgt{\mathrel{\lower2.5pt\vbox{\lineskip=0pt\baselineskip=0pt
           \hbox{$>$}\hbox{$\sim$}}}}
\def\simlt{\mathrel{\lower2.5pt\vbox{\lineskip=0pt\baselineskip=0pt
           \hbox{$<$}\hbox{$\sim$}}}}
\makeatother

\def\mysection#1{{\bf #1.} }

\newcommand{\be}{\begin{equation}}
\newcommand{\ee}{\end{equation}}
\newcommand{\bea}{\begin{eqnarray}}
\newcommand{\eea}{\end{eqnarray}}
\newcommand{\beq}{\begin{eqnarray}}
\newcommand{\eeq}{\end{eqnarray}}

\newcommand{\no}{\nonumber}

\newcommand{\abs}[1]{\left| #1 \right|}

\newcommand{\mHe}{m_\text{He}}

\def\lsim{\mathrel{\rlap{\lower4pt\hbox{\hskip1pt$\sim$}}
     \raise1pt\hbox{$<$}}}         
\def\gsim{\mathrel{\rlap{\lower4pt\hbox{\hskip1pt$\sim$}}
     \raise1pt\hbox{$>$}}}         

\begin{document}

\title{On the Detectability of Light Dark Matter with Superfluid Helium}
\author{Katelin Schutz}
\author{Kathryn M. Zurek}
\affiliation{Theoretical Physics Group, Lawrence Berkeley National Laboratory, Berkeley, CA 94720 \\ Berkeley Center for Theoretical Physics, University of California, Berkeley, CA 94720}

\begin{abstract}
\noindent  
We show that a two-excitation process in superfluid helium, combined with sensitivity to meV energy depositions, can probe  dark matter down to the $\sim$keV warm dark matter mass limit. This mass reach is three orders of magnitude below what can be probed with ordinary nuclear recoils in helium at the same energy resolution. For dark matter lighter than $\sim 100$ keV, the kinematics of the process requires the two athermal excitations to have nearly equal and opposite momentum, potentially providing a built-in coincidence mechanism for controlling backgrounds. 
\end{abstract}

\maketitle

\mysection{Introduction}
\label{sec:intro}
The endeavor to detect and probe dark matter (DM) directly has seen promising strides in recent years, and yet the DM particle remains elusive. Existing nuclear recoil experiments have tightly constrained DM above $m_X$$\, \sim\,$10~GeV \cite{Aprile:2012nq,Akerib:2013tjd,Agnese:2014aze}, a mass scale that is well-motivated by the weakly-interacting massive particle (WIMP) paradigm. WIMPs gained prominence both because of the coincidence of the relic abundance of DM with freeze-out at the weak scale, as well as their connection to the hierarchy problem. 
In recent years, however, broad classes of well-motivated DM  models have emerged with DM candidates having $m_X$$\,<\,$10~GeV \cite{Mohapatra:2000qx,Mohapatra:2001sx,Boehm:2003hm,Strassler:2006im,Pospelov:2007mp,Hooper:2008im,Feng:2008ya,Kaplan:2009ag,Zurek:2013wia,Hochberg:2014dra,Hochberg:2014kqa}. 
In response, new ways of detecting DM via nuclear \cite{Guo:2013dt} and electron recoils have been proposed \cite{Essig:2011nj,Essig:2015cda,Agnese:2013jaa}, and a limit (though still relatively weak) has been set for $m_X \gtrsim 10 \mbox{ MeV}$ \cite{Essig:2012yx}. 

There are two main obstacles to detecting DM down to mass scales as light as the warm DM limit (corresponding to $m_X \sim 1$ keV \cite{Tremaine:1979we,Boyarsky:2008ju,Boyarsky:2008xj}). The first is  that the initial kinetic energy available for scattering, $E_i= \frac{1}{2} m_X v_X^2$, becomes as small as 1 meV for keV-mass DM, with the velocity set by the local velocity dispersion of the Milky Way, $v_X\sim10^{-3}$. The 1 meV scale is well below the energy resolution of current experiments, though substantial technological advances are underway. 
Second, as the DM mass drops below the target mass, momentum conservation enforces that a decreasing fraction of the DM kinetic energy can be transferred to the nucleus or electron in an elastic collision --- the maximum momentum transfer is $q_{max} = 2 m_X v_X$, corresponding to an energy transfer of $2 m_X v_X^2 (m_X/m_T)$, where $m_T$ is the mass of the target electron or nucleus. Thus, as the DM mass decreases, one gets decreasing returns in energy deposition; for instance, keV-mass DM can deposit at most $\sim 10^{-9}$~eV on a light nuclear target like helium. 

One way to bypass these challenges for DM lighter than 1 MeV is to use the target velocity, as was proposed for the detection of keV-mass DM using meV energy depositions on electrons in superconducting aluminum \cite{Hochberg:2015pha,Hochberg:2015fth}. 
Here, we develop an alternative idea for detecting super light dark matter via nuclear interactions. By coupling nuclear DM scattering to {\em multiple} excitations in superfluid helium, we can experimentally probe the whole range of kinematically available DM energy and momentum. 

\mysection{Detection with Superfluid Helium}
To understand why multiple excitations are necessary for probing $m_X\sim$ keV-MeV, first consider a single-phonon process. In superfluid helium, phonons with momentum $k \lesssim 1$ keV (corresponding to wavenumber $k\sim 1\,$\AA$^{-1}$ but in units where $\hbar\,$=$\,c\,$=\,1) have a linear dispersion relation, $\omega= c_s k$, where the sound speed is $c_s\sim10^{-6}$. Since $c_s \ll v_X$, a single on-shell phonon (a phonon obeying the dispersion relation) is unable to absorb an ${\cal O}(1)$ fraction of the DM kinetic energy while still conserving momentum. 
The situation changes dramatically when multi-excitation processes are considered. Two on-shell quasiparticles
 with nearly equal and opposite momentum can absorb all the DM kinetic energy while still conserving momentum. 
There is a phase space suppression  for this configuration, but we will show that for light DM, the constraints from a helium experiment with a kg-year of exposure will complement superconducting aluminum targets. 

As with superconductors, excitations in superfluids do not easily thermalize, making energy deposits detectable above thermal noise. 
We expect that high energy excitations will decay to lower energy (athermal) phonon and roton modes, initiating a shower along the direction of propagation of the initial excitations. In addition, the pair of excitations retains information about the direction of the initial momentum transfer in the DM scattering, though the the size of this effect (and corresponding ability to reconstruct the initial direction of the DM motion) will be suppressed by the small ratio of the momentum transfer to the momentum of the final state excitations. 

Once the energy is deposited in the fluid, it can be measured with transition edge sensors (TESs) or microwave kinetic inductance devices (MKIDs) having the requisite $\sim$meV energy resolution to access the kinetic energy of DM down to $m_X\sim$~1~keV. 
It has been previously argued \cite{Hochberg:2015pha,Hochberg:2015fth} that such an energy resolution could be achieved by shrinking the size and further cooling devices similar to those that have already been designed. Such sensors could equally well be attached to a superconducting aluminum target or to liquid helium, making the development of such sensors highly parallel between the two classes of experiments. We leave a detailed examination of the experimental design for a multi-excitation liquid helium detector to future work \cite{inprep}.

\mysection{Excitations in Superfluid Helium}  
%
%
We begin by using quantum fluid dynamics to parameterize 
second-quantized density and velocity excitations, 
\beq \rho &=  \rho_0 + V^{-\frac{1}{2}} \sum\limits_{k} e^{i (\vec k \cdot \vec r - \omega_k t)} \rho_{\vec k}, \\ \vec{v} &= V^{-\frac{1}{2}}\sum\limits_{k} e^{i (\vec k \cdot \vec r - \omega_k t)} \vec v_{\vec k} \,~~~\eeq
where $V$ is a reference volume and $\rho_0$ is the mean background density.
Free perturbations satisfy the continuity equation $\vec v_{\vec k}= -\vec k\,  \omega_k \rho_{\vec{k}}/\rho_0 k^2$ and the corresponding harmonic oscillator Hamiltonian in Fourier space, \beq H_0& =&  \frac{1}{2} \sum_k \left( \rho_0 v_{\vec k} v_{-\vec k} + \phi_k \rho_{\vec k} \rho_{-\vec k} \right),\eeq 
where $\phi_k$ is the second functional derivative of the energy density with respect to the background density. 
The force constant $\phi_k$ is related to the frequency by $\omega_k^2 = \rho_0 k^2 \phi_k$ and
the frequency of perturbations is given by $\omega_k =  k^2 / 2 \mHe S(k)$. Here $S(k)$ is the static structure factor in units of the mean number density, related to the two-point correlation function of perturbations in the liquid, $\mHe^2 S(k) = \langle \rho_k \rho_{-k}\rangle$. 
This function scales linearly for $k \lesssim 1$~keV giving a linear dispersion relation, and levels off to 1 at high $k\gtrsim5$~keV, giving the typical free-particle dispersion relation \cite{Svensson:1980zz}. 

From the commutation relation between the density and velocity \cite{landau1941theory}, writing $\rho$ and $\vec v$ in terms of the usual creation and annihilation operators, we find \beq \rho_{\vec{k}} &= \mHe \sqrt{S(k)} (a_{\vec k}-a_{-\vec k}^\dagger)\\\vec v_{\vec k} &= -\frac{\vec{k}}{2 \mHe \sqrt{S(k)}} (a_{\vec k}+a_{-\vec k}^\dagger). \eeq
Then, expanding the Hamiltonian to the next (third) order in perturbations, we find, similar to \cite{golub1979storage, golubnotes},
\beq
H_3 = \int d^3 r \left(\frac{1}{2} \vec{v} \cdot \rho \vec{v} + \frac{1}{3!} \frac{\delta \phi_k[\rho_0]}{\delta \rho_0} {\rho}^3\right). 
\label{eq:H3}
\eeq
At small $k$, $\phi_k[\rho_0] = c_s^2/\rho_0$, implying $\delta\phi_k[\rho_0]/\delta\rho_0 = c_s^2(2 u_0 - 1)/\rho_0^2$, where $u_0 \equiv (\rho_0/c_s)(\delta c_s/\delta\rho_0) = 2.84$, as measured by~\cite{abraham1970velocity}. Beyond this regime, the inclusion of the $\rho^3$ term varies between different treatments in the literature and we therefore will drop it for the remainder of this work. We note that this may cause the computed rate to be different by $\mathcal{O}(1)$ factors and will address self-consistent inclusion of the $\rho^3$ term in future work \cite{inprep2}.

This simple picture of quantum fluid perturbations is substantially complicated by the fact that superfluid helium is an interacting Bose fluid. Excitations with a wavelength much larger than the interatomic spacing involves many atoms, implying that a correct description of scattering at low momentum transfer $(q \lesssim 1~$\AA) must include interatomic correlations. Feynman and Cohen \cite{FC} introduced a correction to the ground state wavefunction, ``backflow,'' which accounts for the positions of the other atoms. The method of correlated basis functions (CBF) \cite{Feenberg} is another natural extension of the theory that systematically allows one to compute the response of the fluid to one 
or more excitations. Here we will denote one and two excitation states by $|\vec k \rangle = \rho_{\vec k}^\dagger|0\rangle$ and $|\vec k_1 \vec k_2 \rangle = \rho_{\vec k_1}^\dagger\rho_{\vec k_2}^\dagger|0\rangle$, 
respectively. Due to interactions in the fluid, these states are not orthogonal, $\langle \vec k_1 \vec k_2| \vec k_1 + \vec k_2 \rangle \neq 0$, and they must be orthonormalized. The orthonormalized two-excitation state (denoted with a rounded bracket) is (see for example the discussion in \cite{glyde1994excitations,FG})
\beq
|\vec{k}_1 \vec{k}_2 )=\frac{ \rho_{\vec k_1}\rho_{\vec k_2} - \frac{\langle \vec k_1 + \vec k_2| \vec{k}_1 \vec{k}_2\rangle}{\langle \vec q\,| \vec q\,\rangle}\rho_{\vec k_1 + \vec k_2}}{ \langle \vec{k}_1 \vec{k}_2  | \vec{k}_1 \vec{k}_2  \rangle^{1/2}} \,|0 \rangle .
\eeq
One can then compute the matrix element to create two excitations:
\begin{align}
\label{eq:ME}
( \vec{k}_1 \vec{k}_2| H_3 | \vec q &)  =  -\frac{1}{2 \mHe (S(q) S(k_1) S(k_2))^{1/2}} \times \\ \nonumber &\left( \vec q \cdot \vec k_1 U(k_1) + \vec q \cdot \vec k_2 U(k_2) + q^2 U(k_1) U(k_2) \right),
\end{align}
where 
$U(q) = S(q) - 1$ and where we emphasize again that we are only including the kinetic term in the Hamiltonian \cite{JacksonFeenberg,glyde1994excitations}. 
Results with similar energy and momentum scalings are obtained from the method of collective coordinates \cite{SYK}, as well as in the dielectric formulation \cite{WongGould}. We refer the reader to Ref.~\cite{FG} for a review of these results, and leave a more detailed discussion for future work \cite{inprep2}.

\begin{figure}[t]
\includegraphics[width = 0.42\textwidth]{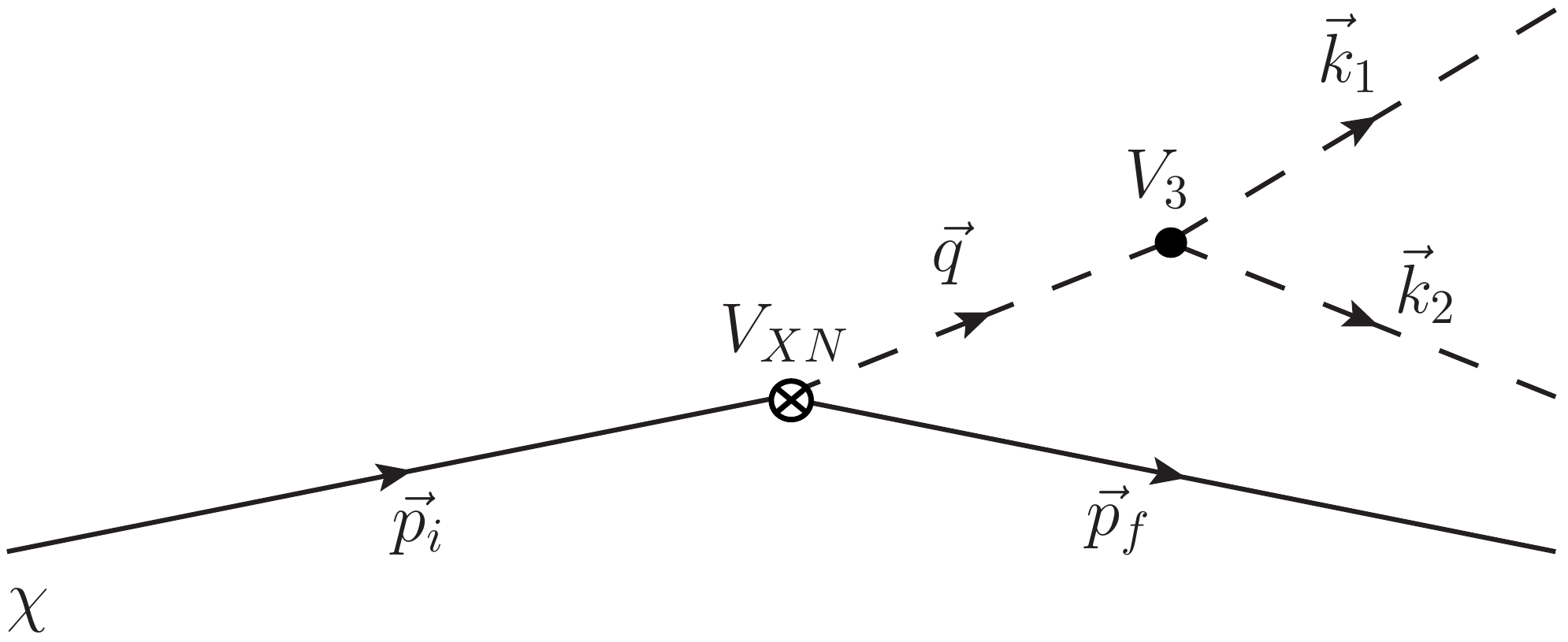}
\caption{The two-excitation process we consider and the corresponding kinematics. The dashed lines denote excitations, while solid lines denote dark matter. \vspace{-0.4cm}}
\label{fig:diagram}
\end{figure}

\mysection{Multi-Excitation Scattering Rates} We now turn to calculating the rate of the interaction shown in Fig.~\ref{fig:diagram}. DM with initial momentum $\vec p_i$ interacts with a helium nucleus initially at rest, transferring momentum $\vec{q}$ and energy $\omega$ to the nucleus. In an ordinary nuclear recoil, the maximum momentum transfer is $q_{\rm max} = 2 m_X v_X$, and a typical energy deposition on the target nucleus $\omega \simeq 10^{-9} \mbox{ eV} (m_X/\mbox{keV})^2$.

As suggested above and depicted in Fig.~\ref{fig:diagram}, more energy can be deposited via nuclear targets when 
energy and momentum $(\omega,\vec q)$ are deposited on a mediating {\em off-shell} excitation. This excitation can come back on shell when the interaction characterized by the Hamiltonian in Eq.~(\ref{eq:ME}) leads to a splitting into two excitations carrying momentum $\vec k_1$ and $\vec k_2$. When $\omega \gg c_s q_{\rm max}$, 
these excitations must be nearly back-to-back in order to conserve momentum. This configuration has suppressed phase space, but we will show that the rate for this process is non-zero. This is also confirmed by the observation of a response in superfluid helium away from the single-excitation dispersion curve (see {\em e.g.} \cite{2016arXiv160502638B} for recent measurements). 

For the practical purpose of predicting DM exclusion constraints that could be achieved with a superfluid helium experiment, we will use the dynamic structure factor $ S(q,\omega)$, defined in relation to the differential scattering rate as 
 \beq
 \frac{d^2\Gamma}{dqd\omega} = \frac{\rho_0 \sigma_N q}{2 m_X \mHe p_i} S(q,\omega).
 \label{eq:Sqomega}
 \eeq
Though we will adopt $S(q, \omega)$ from a recent state-of-the-art numerical simulation \cite{Eckhard}, the remainder of this section will be devoted to deriving the approximate form of $S(q, \omega)$ in order to analytically understand its behavior in the relevant kinematic regimes.

In order to approximate the multi-excitation scattering rate, we first apply Fermi's golden rule, 
\beq
&& \Gamma =  \frac{1}{(2\pi)^{5}} \int  d^3 p_f\, d^3 k_1\, d^3 k_2 \, \frac{\abs{\langle \mathcal{M}\rangle}^2}{16 m_X^2 m_{\rm He}^2} \nonumber\\
&&\times\,  \delta^{(3)}\left(\vec{q}-\vec{k}_1 - \vec{k}_2\right)    \delta\bigg(\omega - \frac{1}{2 \mHe}\bigg(\frac{k_1^2}{ S(k_1) }+ \frac{k_2^2}{S(k_2) }\bigg)\bigg),~~~~
\label{eq:FGR}
\eeq
where the transition rate, following \cite{Reddy:1997yr},  is
$
W_{fi} = \abs{\langle \mathcal{M}\rangle}^2/16 m_X^2 m_{\rm He}^2.
$
 To compute the matrix element $\abs{\langle \mathcal{M}\rangle}^2$, we need the relevant vertices (labeled $V_3$ and $V_{XN}$ in Fig.~\ref{fig:diagram}) and Green's function for the off-shell intermediate state.

We can read off the appropriate matrix element from the Hamiltonian via
$
W_{fi} =  |V_{XN} G(q,\omega) V_3 |^2.
$
Here $V_3 \equiv ( \vec{k}_1 \vec{k}_2| H_3 | \vec q )$ in Eq.~\ref{eq:ME} and $V_{XN} = 2 \pi a \rho(r)/(m_X \mHe)$ in position space, where $a$ is the scattering length (related to the total cross-section by $\sigma_N =4 \pi a^2 $). Meanwhile in momentum space, $V_{XN} = 2\pi a \sqrt{S(q)}/m_X $. We will consider both massive and light mediators such that in momentum space,  $\sigma_N = 16 \pi \alpha_p \alpha_X (f_p Z + f_n (A-Z))^2 m_X^2/(q^2 + m_\phi^2)^2$, where $Z$ is the atomic number, $A$ is the atomic mass, and $m_\phi$ is the mediator mass. The couplings $\alpha_{X,p} = g_{X,p}^2/4\pi$ and $f_{p,n}$ are between the mediator and the DM, proton, and neutron, respectively. The Green's function for momentum transfer in the fluid has the form $G(q, \omega)=(\omega + \mHe c_s^2 + q^2/2\mHe)/(\omega^2 - c_s^2 q^2 - (q^2/2 \mHe)^2)$ \cite{anderson1975helium}; we will proceed under the approximation that $\omega \gg  \mHe c_s^2$ so that the behavior can be approximated as $G(\omega)\sim 1/\omega$.

 \begin{figure*}[t]
\includegraphics[width=0.5\textwidth]{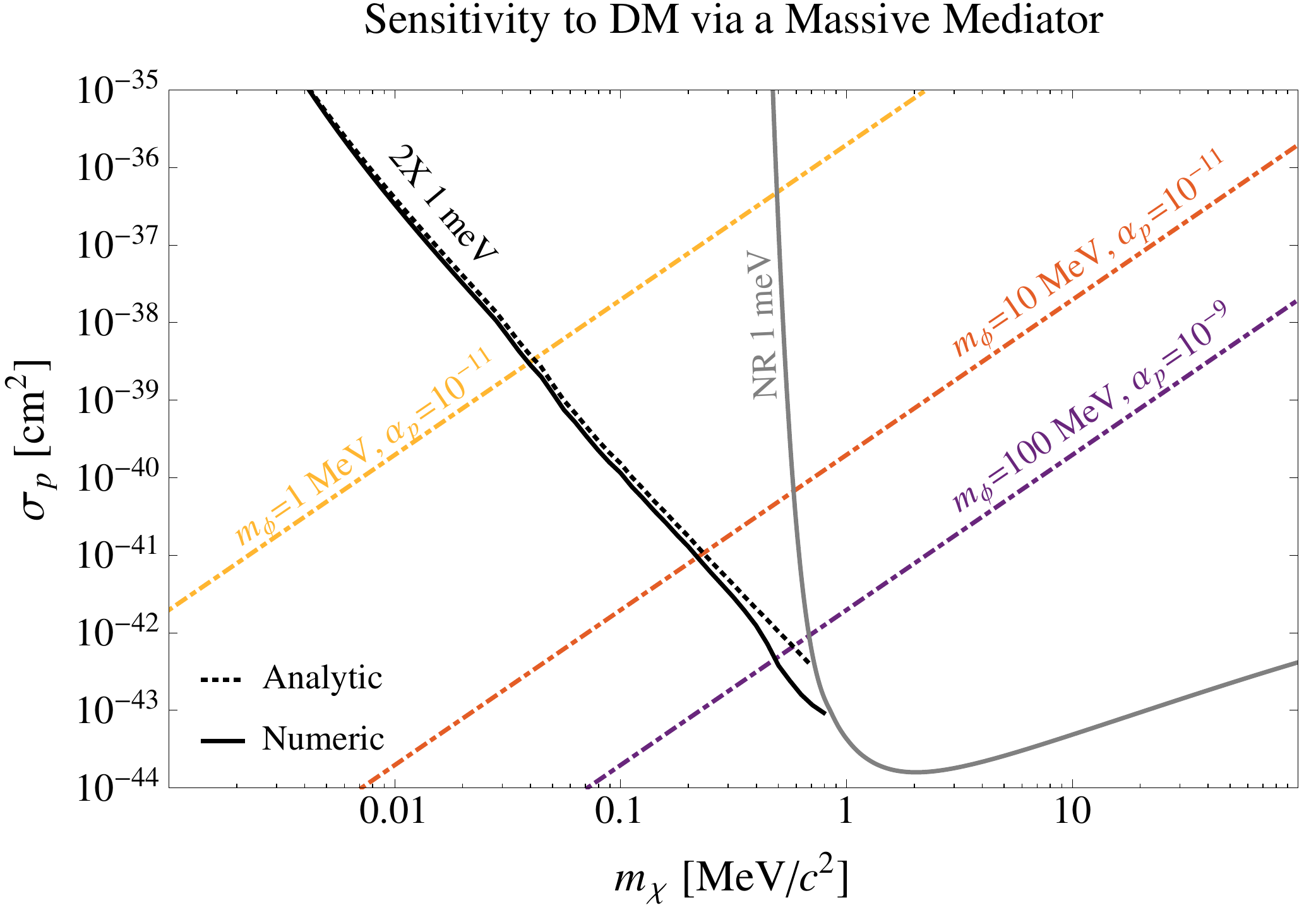}\includegraphics[width=0.5\textwidth]{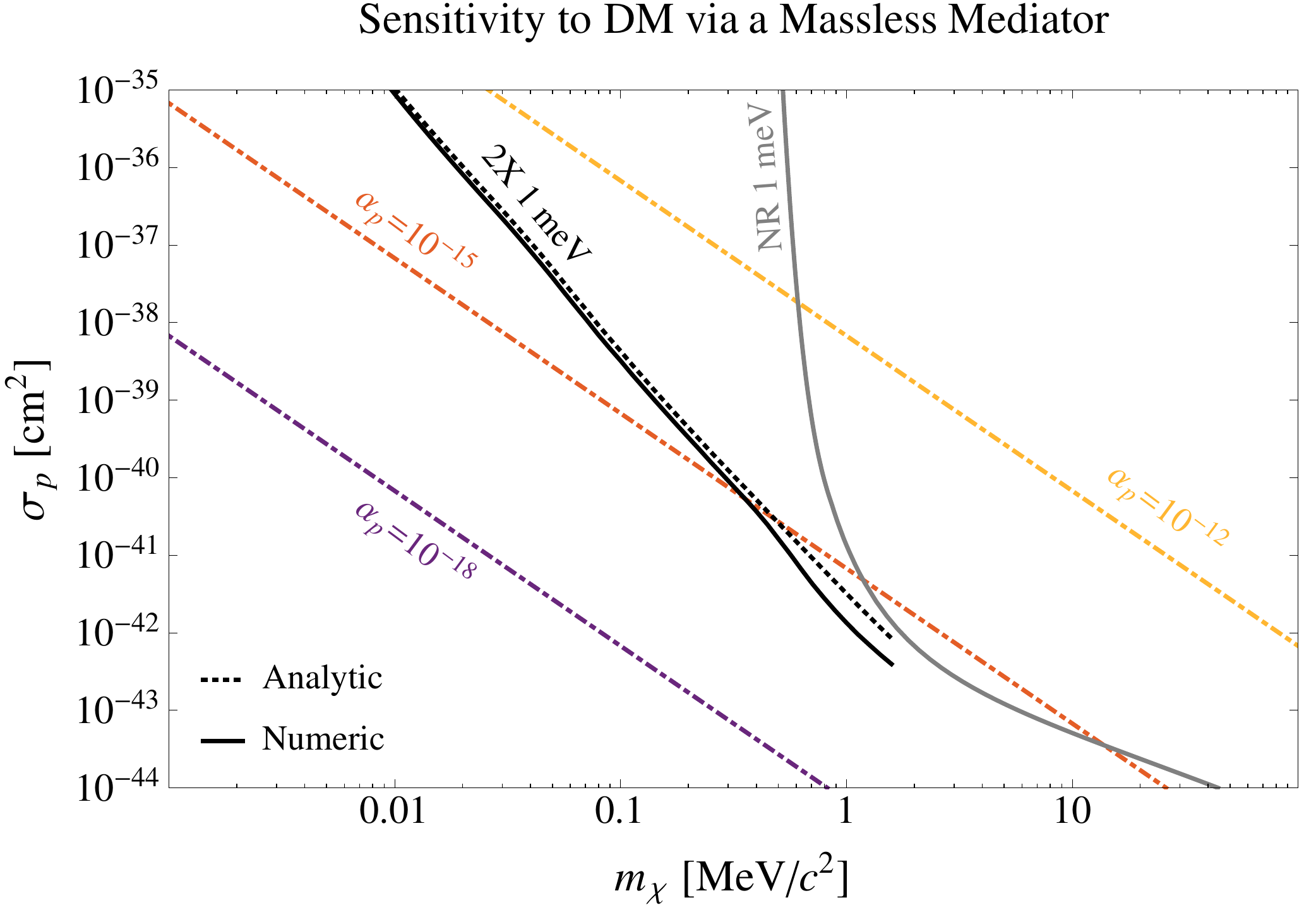}\vspace{-0.2cm}
\caption{95\% confidence level sensitivity expected with a 1 kg-year exposure of superfluid helium. We show both two-excitation processes in the superfluid (labeled 2X) as well as ordinary nuclear recoils (labeled NR), with 1 meV energy resolution in the detectors.  The results are computed analytically via the formula Eq.~\ref{eq:IntRate} (dashed), as well as tabulated from Ref.~\cite{Eckhard} (solid); we have stopped these curves once the scattering begins to probe kinematic regions beyond that tabulated in Ref.~\cite{Eckhard}.   Also shown are benchmarks based on couplings that are consistent with current limits. For the massive mediator, we assume $\alpha_X = 10^{-5}$ for all three curves, while for the light mediator we set $\alpha_X = 10^{-19}$. \vspace{-0.3cm}
}
\label{fig:rate}
\end{figure*}

The rate in Eq.~(\ref{eq:FGR}) can be evaluated for a generic helium dispersion relation and for generic configurations. Here we quote the result in the case that the final state excitations are emitted in a back-to-back configuration, $\vec{k}_1 \approx - \vec{k}_2 \equiv \vec{k}$, which is necessary when ${q} \ll {k}$. This approximation will be good for low-mass DM but will break down by $\sim$1~MeV for energy deposits below 10~meV. 
In order to obtain analytic expressions, we take an approximation for $S(k)$ employed in Ref.~\cite{FG}: $S(k) = k/\sqrt{4 \mHe^2 c_s^2 + k^2}$. While this approximation misses important features such as the roton peak, it does reproduce the correct behavior of $S(k)$ as $k \rightarrow 0$ and when $k \gg 1$~\AA. 

Under these assumptions, the analytic expressions simplify to 
\beq \Gamma =  \int \frac{d^3 k_1 \,d^3 p_f}{8 (2 \pi)^3} \frac{\sigma_N  q^4 h(k)^2 G(\omega)^2}{\mHe^2 m_X^2} \delta \bigg(\omega - \omega_1 - \omega_2 \bigg)~~~~~~
\label{eq:IntRate}
\eeq
where $h(k) =\left( 4 \mHe^2 c_s^2(1 - c^2_\theta) + k^2 -  k \sqrt{k^2+4 \mHe^2 c_s^2}\right)\\/(k^2+4 \mHe^2 c_s^2)$, where $c_\theta$ is the angle between $k_1$ and $q$, and where $\omega_1 = k_1^2/(2 \mHe S(k_1))$ and $\omega_2 = k_2^2/(2 \mHe S(|q-k_2|))$. 
In order to evaluate this further, we note that the $c_\theta$ integral can be carried out utilizing the $\delta$ function, while the $k_1$ integral can be done analytically in the limit that $\omega \gg \mHe c_s^2$. 
We find that 
\beq
S(q,\omega) =  \frac{7 \mHe^{5/2}}{60 \pi^2 \rho_0} \frac{c_s^4 q^4}{\omega^{7/2}},
\label{eq:highomegaSqomega}
\eeq
which is in  in agreement with the $q \rightarrow 0,~\omega\rightarrow \infty$ limit of $S(q,\omega)$ quoted in the literature ({\em e.g.} \cite{FG,WongGould,glyde1994excitations,Griffin}). 
Note that we use the full expression derived from Eq.~\ref{eq:IntRate} in computing rates from the analytic expression; Eq.~\ref{eq:highomegaSqomega} should be considered only a guide to obtain a correct order-of-magnitude estimate at higher DM mass.
Note that due to the steep scaling of $S(q, \omega)$ with $\omega$, the rate will be peaked near the threshold of the detectors, with the rate above $\sim$10 meV energy depositions being negligible.

We will consider two regimes in determining the DM-nucleus cross section: for a heavy mediator  we set $ \sigma_N  =\sigma_p (f_p Z + f_n (A-Z))^2 /f_p^2$, while for a light mediator we set $\sigma_N  =\sigma_p (f_p Z + f_n (A-Z))^2 q_\text{ref}^4/f_p^2 q^4$ where $q$ is in units of a reference momentum at which $\sigma_p$ is evaluated, $q_\text{ref} = v_0 m_X$. In the non-relativistic limit, the final state phase space for the DM is re-written as $d^3 p_f \approx d\omega \,dq\, 2\pi q \,m_X/p_i $. The integral over the momentum transfer is from $p_i - p_f$ to $p_i+p_f$ where in terms of the initial DM energy $E_i = \frac{1}{2}m_X v_X^2$, the momenta are $p_i = \sqrt{2 m_X E_i}$ and $p_f = \sqrt{2m_X(E_i - \omega)}$. We thus obtain for the differential rate, from Eq.~\ref{eq:highomegaSqomega}, 
\beq
\frac{d\Gamma}{d \omega} \simeq \frac{7}{120 \pi^2}\frac{\sigma_p (f_p Z + f_n (A-Z))^2 c_s^4 \mHe^{3/2}}{p_i\omega^{7/2}} \,a(E_i,\omega)~~~~~
\eeq 
where
\begin{align}
a(E_i, \omega)_{m_\phi \gg q} &= \frac{32}{6} m_X^2 \sqrt{E_i (E_i - \omega)} (4 E_i - \omega)(4 E_i - 3 \omega) \no\\
a(E_i, \omega)_{m_\phi \ll q} &= 4 \,q_\text{ref}^4  \sqrt{E_i (E_i - \omega)}.  
\end{align}

\mysection{Detection Rates and Sensitivity Forecasts}  The scattering rate for individual DM particles producing back-to-back excitations can now be converted to a DM detection rate $R$ per target mass via
\begin{equation}
\omega \frac{d R}{d  \omega} = \int d v_X f_{MB}(v_X)\, \omega \,\frac{d \Gamma}{d \omega}\frac{\rho_X}{\rho_0 m_X},
\end{equation}
where $\rho_X$ is the local DM density 0.3 GeV/cm$^3$, $\rho_0$ is the density of liquid helium, and $f_{MB}$ is the Maxwell-Boltzmann distribution of DM in the Milky Way halo,
 \beq f_{MB}(v_X) = \frac{4 \pi v_X^2 e^{-v_X^2/v_0^2} \Theta(v_{esc}- v_X)}{\left(\text{erf}( z) - 2 ze^{- z^2} / \sqrt{\pi}\right) \pi^{3/2} v_0^3} \eeq  with $z = v_{esc}/v_0$ and where $\Theta$ denotes the Heaviside step function. Here we take the root-mean-square velocity $v_0$ to be 220~km/s and the escape velocity $v_{esc}$ to be 500~km/s \cite{Lewin:1995rx}. For both massive and light mediators, the rate is peaked at low $\omega$.

Integrating over deposited energies, in Fig.~\ref{fig:rate} we show the expected sensitivity of a 1 kg-year exposure of superfluid helium to a two-excitation process, assuming a minimum energy sensitivity of 1 meV, and a dynamic range of the sensor up to 10 meV.  We compute the rate utilizing the analytic formula, Eq.~\ref{eq:IntRate} (dashed), as well as tabulated from Ref.~\cite{Eckhard} (solid).  The two are in good agreement, except at masses approaching an MeV, where the tabulated $S(q,\omega)$ also includes a contribution from single phonon emission.  We also show the expected constraints from ordinary nuclear recoils in the fluid, this time allowing for a larger energy deposit (up to 10 eV) in order to capture the full sensitivity to $\sim 100$ MeV dark matter. 
When showing our results, we constrain $\sigma_p$ in the case where $f_p =f_n$, and consider both a light ($m_\phi \ll v_0 m_X$) and heavy ($m_\phi \gg v_0 m_X$) mediator.
The solar neutrino background is small on Helium (see Fig. 3 of \cite{Hochberg:2015fth}), so that the 95\% confidence level from a one-sided Poisson distribution corresponds to 3 events. Other sources of noise can be controlled by the requirement that there be two back-to-back excitations in the final state, though we note that this will be less effective at higher DM masses. As can be seen from the plot, two-excitation processes and nuclear recoils provide highly complementary modes of DM detection, with sensitivity in distinct regions of parameter space. With 1 meV energy resolution TESs, we can therefore employ a single multimodal liquid helium experiment to constrain dark matter masses over five orders of magnitude.

We also show scattering cross-sections corresponding to fixed $\alpha_X,~\alpha_p$ for a given mediator mass. These fixed couplings are chosen to broadly satisfy terrestrial, cosmological and astrophysical constraints. The constraints applied are described in general terms for DM-electron interactions in \cite{Hochberg:2015pha}, and are outlined in great detail in \cite{Hochberg:2015fth}. Existing constraints on DM-nucleon interactions are similar for models of interest here (or in some cases weaker, for instance constraints from big bang nucleosynthesis are weaker in models with DM coupling only to nucleons), so we simply make use of these parameters to emphasize that dark matter models satisfying all terrestrial, astrophysical, and cosmological constraints are within reach of the class of experiments we propose. 
The implications of a superfluid helium experiment for various DM models will be explored in future work \cite{inprep2}. 

\mysection{Conclusions} We have proposed a new method of detecting DM using the quantum fluid dynamics of superfluid helium. With a kg-year exposure, we have demonstrated that a superfluid helium experiment would complement recently-proposed superconductor experiments in detecting low-mass DM scattering on nucleons instead of electrons.
Superfluid helium has additional benefits: (1) the kinematics of the two-excitation process provide a coincidence gate for controlling backgrounds for DM lighter than $\sim$1~MeV, and (2) the same experiment can also search for DM via nuclear recoils off helium nucleii, extending the range of DM masses that can be probed. With the benchmarks outlined in this letter, we anticipate that the ability to probe DM as light as the $\sim$keV warm DM limit will motivate further development of the required technologies for making these ideas experimentally viable. In a separate publication, a specific design will be proposed \cite{inprep}. 

\mysection{Acknowledgments}
We thank Eckhard Krotschek for providing the data of Ref.~\cite{Eckhard} utilized in our numeric estimates, Bob Golub for giving us access to the internal notes to accompany his published paper and for discussions explaining his results, Dan McKinsey for pointing out that the response of liquid helium to multi-phonons/rotons is important, Henry Glyde for a discussion about ultra-cold neutrons in superfluid helium, and Matt Pyle for many conversations about detecting small energy deposits in superconductors and superfluids. We also thank Yonit Hochberg, Michele Papucci, Peter Scherpelz, and Yue Zhao for comments on the manuscript, and Tongyan Lin for collaboration in work to appear \cite{inprep2}. KS is supported by a Hertz Foundation Fellowship and a National Science Foundation Graduate Research Fellowship. KS and KZ are supported by the DOE under contract DE-AC02- 05CH11231.

\bibliography{HeDM}
\end{document}